\documentclass[11pt]{article}

\usepackage{graphicx}
\begin{document}
\oddsidemargin .03in
\evensidemargin 0 true pt
\topmargin -.4in


\def\ra{{\rightarrow}}
\def\a{{\alpha}}
\def\b{{\beta}}
\def\l{{\lambda}}
\def\eps{{\epsilon}}
\def\T{{\Theta}}
\def\t{{\theta}}
\def\co{{\cal O}}
\def\car{{\cal R}}
\def\caf{{\cal F}}
\def\cs{{\Theta_S}}
\def\pr{{\partial}}
\def\tri{{\triangle}}
\def\na{{\nabla }}
\def\S{{\Sigma}}
\def\s{{\sigma}}
\def\sp{\vspace{.15in}}
\def\hs{\hspace{.25in}}

\newcommand{\be}{\begin{equation}} \newcommand{\ee}{\end{equation}}
\newcommand{\bea}{\begin{eqnarray}}\newcommand{\eea}
{\end{eqnarray}}


\begin{titlepage}
\topmargin= -.2in
\textheight 9.5in

\begin{center}
\baselineskip= 18 truept
\vspace{.3in}

\centerline{\Large\bf Non-perturbative quantum effects and stringy degenerate geometries:}
\centerline{\Large\bf Vacuum created pair of $(D{\bar D})_3$-brane by a two form}

\vspace{.6in}
\noindent
{{\bf K. Priyabrat Pandey}, {\bf Abhishek K. Singh}, {\bf Sunita Singh} {\bf and} {{\bf Supriya Kar}\footnote{skkar@physics.du.ac.in }}}

\vspace{.2in}

\noindent

\noindent
{{\Large Department of Physics \& Astrophysics}\\
{\Large University of Delhi, New Delhi 110 007, India}}

\vspace{.2in}

{\today}
\thispagestyle{empty}

\vspace{.6in}
\begin{abstract}
We obtain axionic charged primordial black holes on a vacuum created gravitational pair of $(3{\bar 3})$-brane by the Kalb-Ramond field on a $D_4$-brane in presence of a background open string metric. The new geometries on an emergent pair of $(3{\bar 3})$-brane universe is shown to be influenced by the non-perturbative quantum effects underlying a geometric torsion in a second order formalism. The presence of small extra dimensions 
transverse to the pair in the formalism ensures dynamical scalar fields hidden to a $3$-brane or a ${\bar 3}$-brane universe. We investigate the non-perturbative quantum vacua for their characteristic properties to explain the accelerated expansion of our universe. Interestingly the emergent brane universe is shown to possess its origin in a degenerate stringy de Sitter vacua at an early epoch. A non-perturbative $D_p$-brane world volume correction for $p<3$ is worked out to explain some of the quantum effects underlying a quintessence axion in the string-brane setup. Our analysis reveals that a $D$-instanton can be a potential candidate to source the dark energy in our brane universe.

\baselineskip=14 truept

\vspace{.12in}

\vspace{1in}

\noindent

\noindent

\end{abstract}
\end{center}

\vspace{.2in}

\baselineskip= 16 truept

\vspace{1in}

\end{titlepage}

\baselineskip= 18 truept

\section{Introduction}
Macroscopic black holes are described in Einstein gravity, which are key to the non-linear interactions on a string world-sheet. The string interactions, with a nontrivial background metric, signify the vital role played by a string coupling to incorporate quantum effects in Einstein vacuum. A weak coupling consistently incorporates peturbative geometric (higher genus) corrections to the string world-sheet at tree level and leads to a quantum gravity. The world-sheet conformal invariance at the quantum level ensures an effective space-time string action in ten dimensions, which is known to incorporate higher order $\alpha'$ (string slope parameter) quantum corrections to Einstein vacuum. Generically the low energy string effective actions have been investigated in presence of the Neveu-Schwarz (NS) background \cite{candelasHS,freed,callan} and they have given birth to $(3+1)$-dimensional black holes in string theory \cite{garfinkle,giddings-strominger}. In addition the string black holes in various dimensions have been investigated using vector and chiral gauged WZW models, which possess an exact CFT description \cite{witten-2DBH,n-ishibashi,kar-kumar,kar-khastgir-sengupta,nappi-witten}. In other words, a perturbative string theory enumerates the quantum phenomena to the $(3+1)$-dimensional Einstein vacuum by consistently incorporating extra small spatial dimensions at Planck scale. An extra dimension to Einstein vacuum is believed to unfold the mysteries of dark energy in our universe.

\sp
\noindent
On the other hand, the quantum effects to Einstein gravity may well be described by a non-perturbative world underlying a strong-weak coupling duality in ten dimensional superstring theories \cite{ashoke-sen-ijmpa}. A non-perturbative quantum effect is believed to be sourced by compactified extra space dimension(s) to various stringy vacua. Thus their essence may be described by the hidden Liouville scalar(s) dynamics to our $(3+1)$-dimensional universe. In particular, the type IIA superstring theory in a strong coupling limit is known to incorporate an extra spatial dimension on $S^1$ and has been identified with the non-perturbative $M$-theory in eleven dimensions. Generically $M$-theory has been shown to be identified with the stringy vacua in various dimensions \cite{witten-M}. In a low energy limit $M$-theory is known to describe an eleven dimensional supergravity. However a complete non-perturbative formulation of M-theory is not known. Past attempts to formulate $M$-theory to describe the strong coupling regime in type IIA and type IIB superstring theories were based on matrix theories \cite{BFSS,IKKT}. Another intriguing tool was AdS/CFT duality, which successfully explores the non-perturbative world of gauge and gravity theories \cite{maldacena,witten-ads}.

\sp 
\noindent
In the context, Dirichlet $(D)$ p-brane in type IIA or IIB superstring theory is believed to be a potential candidate to describe a non-perturbative world due to their Ramond-Ramond (RR) charges \cite{polchinski}. In addition, a $D$-brane is known to form a stable bound state with the fundamental string
\cite{witten-bound}. The boundary dynamics of an open string has also been explored using a path integral formalism to describe the interaction of a D-brane with the fundamental string \cite{kar-kazama,kar-npb,kar-npb2,kar-ijmpa}. Analysis have revealed geometrical significance of the background NS field which sources a torsion in space-time \cite{candelasHS,freed}. For a constant NS field, the open string boundary dynamics in the world-volume of a D-brane was remarkably shown to govern a non-commutative $U(1)$ gauge theory, which was equivalently described by a non-linear $U(1)$ gauge theory \cite{seiberg-witten}. Subsequently, various near horizon black hole geometries \cite{gibbons,mars,kar-panda-EM,ishibashi,kar-majumdar,kar-majumdar2,kar-majumdar3,kar-ds,kar-2,liu,zhang2,kar-3} 
have been explored using an open string effective metric sourced by a constant NS field on a $D$-brane. Nevertheless, the mathematical simplicity does not allow a non-constant NS field to couple to the open string boundary dynamics. In principle, a generic NS field is believed to incorporate a new geometry with a varying non-commutative parameter on a $D$-brane. Thus a non-constant NS field shall associate closed string modes to a BPS $D$-brane and hence the set-up may appropriately be described by a non-BPS brane. 

\sp
\noindent
In the recent past, a string-brane set-up underlying such a non-BPS brane configuration was constructed by the authors \cite{abhishek-JHEP,abhishek-PRD,abhishek-NPB-P}. It was shown that the non-BPS brane may well be described by a non-perturbative phenomenon 
leading to a vacuum created gravitational pair of $(D{\bar D})_3$-brane by the Kalb-Ramond (KR) quanta on a $D_4$-brane at the cosmological horizon
of a background (open string) de Sitter geometry. Interestingly, the absorption of the KR field in the five dimensional $U(1)$ gauge theory modifies the covariant derivative, which in turn was shown to describe an effective torsion curvature in a second order formalism. 

\sp
\noindent
Alternately, the local degrees in KR field have been transformed to project a dynamical NS field on a vacuum created non-BPS pair of $(D{\bar D})_3$-brane presumably underlying an early phase of our universe. The momentum conservation, in the string-brane set-up, ensures that the vacuum created gravitational 3-brane and anti 3-brane cannot annihilate each other and defines a stable pair of $(3+1)$-dimensional brane/anti-brane universe. In fact, the cosmological creation of a pair of 
non-BPS $(D{\bar D})_9$-brane in a different context was discussed by Majumdar and Davis \cite{majumdar-davis}. Their work was primarily motivated to address the strong coupling regime in type IIA superstring leading to the $M$-theory. A brief out-line to exploit the strength of our string-brane 
set-up or model for an enhanced understanding of the M-theory has been attempted by one of the authors \cite{kar-JAAT}.

\sp
\noindent
It is worth mentioning that a pair of brane and anti-brane is known to break the supersymmetry and hence is described by a non-BPS brane. Past analysis has revealed that some of the non-BPS states are stable in superstring theories \cite{Ashoke-Sen}. It was shown that the (open string) tachyon condenses on a non-BPS brane to describe a stable pair of $(D{\bar D})$-brane \cite{Ashoke-Sen-P}. In fact the energy spectrum underlying a pair corresponds to the lightest states, which are defined with a set of charges. They are stable because the total charge is conserved. A pair of brane/anti-brane has also been analyzed at a finite temperature to establish its stability \cite{kenji-hotta}. In fact, superstring theory is known to describe a large number of non-perturbative objects which may be described as the exotic branes \cite{boer-shigemori}.

\sp
\noindent
Along the theme, Schwinger pair production mechanism is known to be instrumental to gain insights on diverse quantum phenomenon in gravitation and cosmology \cite{schwinger}. The novel idea possesses its origin in quantum field theory which elegantly describes a (positron and electron) pair production at a vacuum by an electromagnetic quanta. The pair production tool was exploited to explain the Hawking radiation phenomenon \cite{hawking}, where an incident photon generates a pair of charged particle and anti-particle at the event horizon of a black hole. Furthermore the mechanism may seen to generalize to produce a pair of charged cosmic string and anti-string at an early era. The emergent pair is governed by the Kibble mechanism \cite{kibble}, which sources higher dimensional topological defects and is manifested via a series of phase transitions from a high temperature phase to a low temperature. In particular the symmetry at a higher temperature is spontaneously broken to yield the topological defects at a lower temperature. Furthermore, the Schwinger effect has been exploited in string theory to investigate a pair creation of open strings in an electric field \cite{burgess,bachas-porrati,bachas}. Importantly the pair production mechanism has been extended to obtain a cosmological creation of a 
pair of $(D{\bar D})$-brane in superstring theory \cite{majumdar-davis,majumdar-davis-2} and the open string pair creation from the world-sheet instanton \cite{schubert-torrielli}. 

\sp
\noindent
In the past, the vacuum created pair of $(D{\bar D})$-brane was explored to investigate the cosmological behaviour of our universe. It was inspired by a fact that extra spatial dimensions with small radii of compactifications are believed to play a significant role in the study of quantum cosmology. It urges to review the recent developments in string theory with a renewed interest. In particular we have obtained emergent Kerr-brane(anti-brane) geometries, which have been shown to describe the Einstein vacuum in a low energy \cite{sunita-NPB,sunita-IJMPA}. A generalization underlying a CFT on a $D_5$-brane has been shown to describe Schwarzschild de Sitter and topological de Sitter on a $(D{\bar D})_4$-brane by one of the authors (SK) in collaborations \cite{richa-IJMPD}. Generically the brane-world models have been successful to address some of the cosmological puzzles such as the inflation \cite{Shiu-tye-BraneInflation,Shiu-tye-BraneInflation2,Dvali-D-inflation,majumdar-Dinflation,alex-inflation-DD,SCDavis-braneCosmology}. For a review, see ref.\cite{quevedo-lecture}. 

\sp
\noindent
In addition, the issues pertaining to the cosmological inflation has been addressed by exploiting a cascade of fluxes
with a pair of brane/anti-brane \cite{Amico-Gobbetti-Kleban-Schillo,Amico-Gobbetti-Kleban-Schillo-2}. The extra dimensions, to a vacuum created pair of a lower dimensional $(D{\bar D})$-brane, are identified with the transverse scalar fields and possess their origin in a gauge field on a space filling $D_9$-brane. Thus an extra dimension, in the disguise of a scalar field, sources a quintessence in $(3+1)$-dimensional Einstein gravity. The quintessence is believed to be potential candidate to describe the dark energy in our universe. For instance, some of the relevant computations pertaining to a quintessence scalar dynamics in higher dimensional gravitation and cosmology have been worked out in refs.\cite{chen-jing-quintessence,chen-wang-su-quintessence,chen-pan-jing,huan-jun-quintessence}. 

\sp
\noindent
Interestingly our string-brane set-up has been exploited to describe the inflationary cosmology underlying a rapid accelerated expansion of our universe \cite{priyabrat-EPJC,priyabrat-IJMPA}. An extra fifth dimension transverse to both the four dimensional world-volumes has been argued to play a significant role to describe a nontrivial vacuum geometries. In fact, a large extra dimension incorporates the low energy closed string modes in between a brane and an anti-brane. It was shown that the emergent geometry is described by a low energy effective string vacuum in presence of a non-perturbative ($D_3$-brane or an anti $D_3$-brane) quantum correction. 
The torsion condenses in a low energy limit and hence the reduced geometry becomes Riemannian \cite{kpss-ws1,kpss-ws2,pssk-JAAT}.

\sp
\noindent
In the paper, we focus on new quantum non-perturbative geometries in presence of a propagating torsion on a vacuum created pair of gravitational $(3{\bar 3})$-brane. They are indeed a new class of quantum vacua as they are shown to be flat in a low energy limit, $i.e.$ in absence of a geometric torsion. It would imply an open string propagation in a Minkowski background. It confirms that the new quantum geometries are exact in a perturbative string theory. They donot possess any classical geometric (Einstein) vacuum. Thus, the emergent quantum vacua, obtained in this paper, are indeed different from the brane/anti-brane quantum geometries previously obtained in our other papers \cite{kpss-ws1,kpss-ws2,abhishek-JHEP,abhishek-PRD,abhishek-NPB-P,pssk-JAAT,kar-JAAT,sunita-NPB,sunita-IJMPA,priyabrat-EPJC,priyabrat-IJMPA,richa-IJMPD}.

\sp
\noindent
Furthermore, the brane geometries in this paper are shown to be sourced by an axionic scalar underlying a vacuum created gravitational pair of $(3{\bar 3})$-brane. The emergent brane geometries are shown to degenerate. Their degeneracy is sourced by a string charge underlying a NS field in the string bulk. In principle, the degeneracy may seen to be sourced by a lower form obtained from a higher form on $S^1$ or on $S^1\times S^1$. The degenerate brane vacua at string scale may intuitively be identified with the notion of lower dimensional $D$-brane within a higher dimensional brane \cite{douglas-branes}. 

\sp
\noindent
We work out the quantum effects and show that the low energy perturbative string vacuum receives a non-perturbative $D$-brane world-volume correction in the string-brane model. The axionic sourced non-perturbative quantum geometries 
are further analyzed for two larger length scales to qualitatively describe a laboratory black hole in a medium energy regime and a semi-classical brane universe in a moderately low energy regime. Analysis under Weyl scaling(s) are performed to explore some of the effective (Anti) de Sitter brane vacua in presence of non-perturbative correction(s). A quantum correction is shown to be approximated by a varying energy density underlying an axionic quintessence. It is argued that a $D$-instanton, sourced by an axion, presumably incorporates the quintessential effects to our universe. These non-perturbative quantum effects indeed provide clue for extra dimension(s) to our $(3+1)$-dimensional universe.

\sp
\noindent
We plan the paper as follows. We begin with a set-up underlying an effective curvature on a $D_4$-brane in presence of a background open string metric in section 2. A gauge choice and various plausible torsion geometries on a vacuum created gravitational pair of $(3{\bar 3})$-brane are obtained in section 3. Non-perturbative quantum effects underlying a lower dimensional $D$-brane are explicitly worked out in section 4. The brane geometries are Weyl scaled appropriately in section 5 to obtain a de Sitter brane and an AdS brane in the formalism. The brane universe is shown to be describe by a quintessence axion. We conclude with a brief summary in section 5.

\section{Preliminaries}
\subsection{Geometric torsion ${\mathbf{\cal H}_3}$ in ${\mathbf{5D}}$}
To illustrate the emergent scenario, we begin with a KR field $U(1)$ dynamics, in presence of a background (open) string metric, on a $D_4$-brane. Needless to mention that a nontrivial background metric in the world-volume dynamics is sourced by a constant NS field in string theory. It may be noted that the set-up evolves with two independent two forms. One of it is dynamical KR field and the other describes a constant NS field. An appropriate covariant derivative is explored in presence of a world-volume torsion $H=dB$ connection, which in turn incorporates a geometric torsion ${\cal H}_3$ in a second order curvature formalism \cite{kpss-ws1,kpss-ws2,pssk-JAAT,kar-JAAT}. The presence of an axion has been explored to address a quintessence scalar dynamics hidden to a vacuum created gravitational $3$-brane universe within a pair.

\sp
\noindent
On the other hand, a $D_4$-brane is a dynamical object in ten dimensional type IIA superstring theory. Its five dimensional world-volume dynamics may be approximated by a Dirac-Born-Infeld (DBI) action, which describes a non-linear one form ${\cal A}_{\mu}$ in a $U(1)$ gauge theory. For a constant background metric $g_{\mu\nu}$, the dynamics of ${\cal A}_{\mu}$ leads to a non-linear field strength ${\bar{F}}_{\mu\nu}=(2\pi\alpha')F_{\mu\nu}
= \left ({\bar F}_{\mu\nu}^{\rm linear} + B_{\mu\nu}^{NS}\right )$. It is given by
\be
S= -{1\over{4C_1^2}}\int d^5x\ {\sqrt{-\det \left ( g +  {\bar{F}}\right )_{\mu\nu}}} \; ,\label{gauge-1}
\ee 
where $C_1^2=(4\pi^2g_s){\alpha'}^{5/2}$. Alternately, the Poincare dual to the field strength on a $D_4$-brane underlie a dynamical KR field  $B_{\mu\nu}$ and is known to incorporate a gauge theoretic torsion $H_3=dB_2$. The KR field dynamics, in presence of a background open string metric 
${\tilde G^{(NS)}}_{\mu\nu}=\left ( g_{\mu\nu} - B^{(NS)}_{\mu\lambda}{B^{\lambda (NS)}}_{\nu}\right )$, is given by
\be
S=- {1\over{12C_2^2}}\int d^5x\ {\sqrt{-\det {\tilde G}^{({\rm NS})}}}\ H_{\mu\nu\lambda}H^{\mu\nu\lambda}\ ,\qquad {\rm where}\quad C_2^2=(8\pi^3g_s){\alpha'}^{3/2}\ .\label{gauge-2}
\ee
The significance of the constant NS field in addition to the local modes of the KR field have been addressed in an effective curvature formalism on a $D_4$-brane \cite{kpss-ws1,kpss-ws2,abhishek-JHEP,abhishek-PRD,pssk-JAAT}. A priori, a gauge theoretic torsion $H_3$ has been identified with a connection, which in turn modifies the covariant derivative to ${\cal D}_{\mu}$ on a $D_4$-brane. The mathematical construction implies an absorption of KR field quanta and hence the ${\cal D}_{\mu}$ describes a ``fat'' brane or a string-brane model. It has been argued to describe a
a vacuum created pair of gravitational $(3{\bar 3})$-brane underlying a geometric torsion curvature in the string-brane model. The gauge connections are appropriately coupled to the NS field and has been shown to define a geometric torsion ${\cal H}_3$. Generically the geometric torsion for an arbitrary two form takes an usual form:
\be
{\cal H}_{\mu\nu\lambda}= {\cal D}_{\mu}B_{\nu\lambda}\ +\;  {\rm cyclic\ in}\; (\mu, \nu, \lambda )\ ,\label{gauge-3}
\ee  
\be
{\rm where}\qquad {\cal D}_{\lambda}B_{\mu\nu}=\nabla_{\lambda}B_{\mu\nu} + {1\over2}H_{{\lambda\mu}}^{\rho}B_{\rho\nu} - {1\over2} H_{{\lambda\nu}}^{\rho}B_{\rho\mu}\ .\label{gauge-4}
\ee
An iterative incorporation of the constant NS field correction in the covariant derivative leads to an exact derivative in a perturbative gauge theory. It may seen to define a non-perturbative covariant derivative in a second order formalism underlying a geometric realization. Then, the geometric torsion may be given by
\bea
{\cal H}_{\mu\nu\lambda}&=&\nabla_{\lambda}B_{\mu\nu} + {1\over2}{{{\cal H}}_{\lambda\mu}}^{\rho}B_{\rho\nu} - {1\over2}{{\cal H}_{\lambda\nu}}^{\rho}B_{\rho\mu}\ + {\rm cyclic\ in}\; (\mu, \nu, \lambda)\ ,\nonumber\\
&=&H_{\mu\nu\lambda} + 3{{\cal H}_{[\mu\nu}}^{\alpha}{B^{\beta}}_{\lambda ]}\ g_{\alpha\beta}\nonumber\\
&=&H_{\mu\nu\lambda} + \left ( H_{\mu\nu\alpha}{B^{\alpha}}_{\lambda} + \rm{cyclic\; in\;} \mu,\nu,\lambda \right )\ +\ H_{\mu\nu\beta} {B^{\beta}}_{\alpha} {B^{\alpha}}_{\lambda}\ +\ \dots \; . {}\label{gauge-5}
\eea
Since all the local degrees in KR field have been transformed to the constant NS field, $i.e.\ \nabla_{\lambda}B^{{\rm NS})}_{\mu\nu}=0$, the geometric torsion should be viewed via a dynamical NS field. As a result, the brane set-up in presence of an open string background metric may be identified as a string-brane model. Then, the geometric torsion in the string-brane model may precisely be defined as: 
\bea
{\cal H}_{\mu\nu\lambda}&\rightarrow&{1\over2}{{{\cal H}}_{\lambda\mu}}^{\rho}B^{(NS)}_{\rho\nu} - {1\over2}{{\cal H}_{\lambda\nu}}^{\rho}B^{(NS)}_{\rho\mu}\ + {\rm cyclic\ in}\; (\mu, \nu, \lambda)\ ,\nonumber\\
&=&3{{\cal H}_{[\mu\nu}}^{\alpha}{B^{(NS)\beta}}_{\lambda ]}\ g_{\alpha\beta}\nonumber\\
&=&\left ( H_{\mu\nu\alpha}{B^{(NS)\alpha}}_{\lambda} + \rm{cyclic\; in\;} \mu,\nu,\lambda \right )\ +\ H_{\mu\nu\beta} {B^{(NS)\beta}}_{\alpha} {B^{(NS)\alpha}}_{\lambda}\ +\ \dots \; . {}\label{gauge-55}
\eea
It shows that the geometric torsion primarily take into account the dynamics of the NS field leading to a torsion $H$ in the perturbative string effective action. A dynamical NS field on a ``fat'' $D_4$-brane breaks the supersymmetry and leads to a non-BPS brane configuration, $i.e.$ a 
pair of gravitational $(3{\bar 3}$-brane. The presence of string bulk due to the dynamical NS field ensures the closed string modes coupling to the otherwise BPS brane. Importantly, the geometric torsion in the string-brane model is exact and hence is non-perturbative. In addition, the $U(1)$ gauge invariance of a geometric torsion under a two form transformation shall explicitly be realized in an effective space-time curvature formalism. Nevertheless, in the modified $U(1)$ gauge theory, a constant NS field acts as a perturbation parameter and is distinctly placed than the propagating KR field. Thus, the gauge invariance $\delta_{\rm nz}{\cal H}_3=0$ may be retained by the propagating modes of KR field as: $\delta_{\rm nz} H_3=0$ and $\delta_{\rm nz}B^{\rm global}=0$.

\subsection{Space-time curvature on a ${\mathbf{D_4}}$-brane}
The first order KR field dynamics on a $D_4$-brane in presence of a background open string metric may lead to an emergent geometric torsion dynamics 
in $(3+1)$-dimensions in a second order formalism \cite{abhishek-JHEP}. A priori, the emergent torsion curvature was shown to be governed by a
a fourth order generalized tensor underlying a vacuum created gravitational pair of $(3{\bar 3})$-brane \cite{abhishek-PRD}. The reducible curvature takes a  form:
\be
4{{\tilde{\cal K}}_{\mu\nu\lambda}{}}^{\rho}= 2\partial_{\mu}{{\cal H}_{\nu\lambda}}^{\rho} -2\partial_{\nu} {{\cal H}_{\mu\lambda}}^{\rho} + {{\cal H}_{\mu\lambda}}^{\sigma}{{\cal H}_{\nu\sigma}}^{\rho}-{{\cal H}_{\nu\lambda}}^{\sigma}{{\cal H}_{\mu\sigma}}^{\rho}.\label{gauge-6}
\ee
The generalized tensor is antisymmetric under an exchange of indices within a pair and is not symmetric under an exchange of its first pair of indices with the second. It differs from the Riemannian tensor $R_{\mu\nu\lambda\rho}$. However for a non-propagating torsion 
${\tilde {\cal K}}_{\mu\nu\lambda\rho}\rightarrow R_{\mu\nu\lambda\rho}$. The second order curvature tensor and the irreducible curvature scalar are worked out to yield: 
\bea
4{\tilde{\cal K}}_{\mu\nu}&=& -\left (2\partial_{\lambda}{{\cal H}_{\mu\nu}}^{\lambda} +
{{\cal H}_{\mu\rho}}^{\lambda}{{\cal H}_{\lambda\nu}}^{\rho}\right )
\nonumber\\
{\rm and}\qquad {\tilde{\cal K}}&=& -{1\over{4}}{\cal H}_{\mu\nu\lambda}{\cal H}^{\mu\nu\lambda}
\ .\label{gauge-7}
\eea
As a result, the effective torsion curvature constructed in a string-brane set-up describes a non-perturbative curvature tensor.
In a second order formalism, an effective ${\cal H}_3$ dynamics underlying a vacuum created pair of $(3{\bar 3})$-brane may be approximated by
\be
S_{\rm {3-{\bar 3}}}= {1\over{3C_2^2}}\int d^5x {\sqrt{-\det {\tilde G^{(NS)}}_{\mu\nu}}}\;\;\ {\tilde{\cal K}}\ .\label{gauge-6}
\ee
Generically, the open string metric may be re-expressed in terms of the Poincare dual field strength. It takes a form: 
\be
{\tilde G}_{\mu\nu}=\left ( {{\tilde G^{(NS)}}_{\mu\nu}} + C\ {\bar{\cal H}}_{\mu\lambda\rho}{{\cal H}^{\lambda\rho}}_{\nu}\right )\ ,\label{gauge-7}
\ee
where $B_{\mu\nu}^{(NS)}$ signifies a constant NS field and appropriately couples to an electromagnetic field $F_{\mu\nu}^{\rm linear}$ to retain the $U(1)$ gauge invariance. However the underlying $U(1)$ gauge invariance of the curvature scalar ${\tilde{\cal K}}$ apparently incorporate a condition:
\be
C\ (2\pi\alpha'){H_{\alpha\beta}}^{\mu}\ {{\cal H}^{\alpha\beta\nu}} = f^{\mu\nu}_q\ .\label{gauge-9}
\ee
Nevertheless, the perturbation gauge theory re-assures an infinitesimally small constant (two form) mode coupling in a geometric torsion (\ref{gauge-5}) and hence the condition may be ignored. Then, the emerging notion of a metric underlying a propagating geometric torsion in the non-perturbative curvature model is given by
\be
f_{\mu\nu}^q\ =\ C\ {\cal H}_{\mu\alpha\beta} {{\bar{\cal H}}^{\alpha\beta}}_{\nu}\ .\label{gauge-10}
\ee
The symmetricity under an interchange $\mu\leftrightarrow \nu$ in the r.h.s. ensures that the naive condition on a two form becomes insignificant. Interestingly, the emergent metric incorporates the notion of space-time in the curvature scalar ${\tilde{\cal K}}$ and hence justifies the open string effective metric arguably obtained in eq.(\ref{gauge-7}) on a $D_4$-brane. The emerging notion of a metric $f^q_{\mu\nu}$ ensures a space-time curvature scalar ${\tilde{\cal K}}$ underlying a non-perturbative set-up. It incorporates a non-perturbative correction to the open string metric ${\tilde G}_{\mu\nu}$ established on a $D$-brane \cite{seiberg-witten}. In other words, a propagating geometric torsion is responsible for the emergent metric fluctuation $f^q_{\mu\nu}$. Thus the dynamical effects, of a geometric torsion, appear to modify the Riemannian curvature tensor ${R_{\mu\nu\lambda}}^{\rho}$ to an effective curvature tensor ${{\tilde{\cal K}}_{\mu\nu\lambda}}^{\rho}$. 

\sp
\noindent
In a gauge choice for a non-propagating torsion, the effective curvature ${{\tilde{\cal K}}_{\mu\nu\lambda}}^{\rho}$ reduces to ${R_{\mu\nu\lambda}}^{\rho}$. In the case the emergent scenario on a brane universe is described with a large extra dimension. Then a $D$-brane or an anti $D$-brane correction, sourced by a geometric torsion, may be ignored to describe Einstein vacuum in the string-brane model. Furthermore various brane geometries obtained in the model ensure that a geometric torsion contribution to the emergent metric components are generically higher order in $(1/r)$ and hence they play a significant role for small $r$ underlying a microscopic brane universe \cite{abhishek-JHEP,abhishek-PRD,abhishek-NPB-P,sunita-NPB,sunita-IJMPA}. It provokes thought to believe that the metric fluctuations (\ref{gauge-10}) signify a quantum correction to a (perturbative) low energy closed string vacuum.

\sp
\noindent 
Furthermore, the generalized curvature dynamics on $S^1$ may be analyzed to describe a vacuum created gravitational pair of $(3{\bar 3})$-brane in the formalism, Intuitively, the gravitational brane and anti-brane universes are approximated by two independent curvatures in $4D$. They are given by
\be
S_{(3-{\bar 3})}^{\rm eff}= {1\over{3\kappa^2}}\int d^4x {\sqrt{-{G^{(NS)}}}}\ \left ( {\cal K}\ -\ {3\over4} {\bar{\cal F}}_{\mu\nu}
{\cal F}^{\mu\nu} \right )\ .\label{gauge-7}
\ee
\bea
{\rm where }\qquad {\cal F}_{\mu\nu}&=&{\cal D}_{\mu}A_{\nu}-{\cal D}_{\nu}A_{\mu}\ ,\nonumber\\
&=&F_{\mu\nu}\ +\ {{\cal H}_{\mu\nu}}^{\alpha}A_{\alpha}\ .\label{T-23}
\eea
The curvature scalar ${\cal K}$ is essentially sourced by a dynamical two form in a first order formalism. Its field strength is appropriately modified $H_3\rightarrow {\cal H}_3$ to describe a propagating torsion in four dimensions underlying a second order formalism. The fact that a torsion is dual to an axion on a gravitational $3$-brane, ensures one degree of freedom. In addition ${\cal F}_{\mu\nu}$ describes a geometric one form field with two local degrees on a gravitational $3$-brane. A precise match among the (three) local degrees of torsion in ${\tilde{\cal K}}$ on $S^1$ with that in ${\cal K}$ and ${\cal F}_{\mu\nu}$ reassure the absence of a dynamical dilaton field in the frame-work. The result is consistent with the fact that a two form on $S^1$ does not generate a dilaton field. Alternately an emergent metric tensor ${\tilde G}_{\mu\nu}$, essentially sourced by a dynamical two form, does not formally evolve under a compactification of an underlying $U(1)$ gauge theory (\ref{gauge-6}) on $S^1$. The background metric $g_{\mu\nu}$, being non-dynamical on a $D_4$-brane, can not generate a dynamical scalar field or dilaton when compactified on $S^1$ in the frame-work.

\section{Degenerate vacua}

\subsection{Gauge choice leading to torsion geometries in 4D}
A geometric torsion, underlying a $U(1)$ gauge theory, on a $D_4$-brane may alternately be viewed via a vacuum created gravitational pair of $(3{\bar 3})$-brane. Within a pair, the world volumes are distinctly separated by an extra fifth dimension. Thus an irreducible scalar curvature ${\tilde{\cal K}}$ on $S^1$ generically reduces to ${\cal K}$ on a gravitational $3$-brane and a geometric ${\cal F}_2$ on an anti $3$-brane or vice-versa. An ansatz for the gauge fields on a $3$-brane is worked out to yield:
\bea
&&B_{t\phi}=\ -P \cos\theta\nonumber\\
{\rm and}&&A_t=\ {1\over{\sqrt{2\pi\alpha'}}}\left ( Q_1 \cos\left [ \frac{P{\sqrt{2\pi\alpha'}}}{r}\right ]\ -\ Q_2 \sin\left [\frac{P{\sqrt{2\pi\alpha'}}}{r}\right ]\right )\ ,\label{T-1}
\eea
where $P$ and $(Q_1,Q_2)$ signify the non-linear charges respectively under the NS field and the gauge field on a vacuum created pair of $(3{\bar 3})$-brane. The gauge choice incorporates a geometric torsion which is given by
\be
{\cal H}_{t\theta\phi}\ \rightarrow\  H_{t\theta\phi}=  {{-P}\over{\sqrt{2\pi\alpha'}}}\sin\theta\ .\label{T-2}
\ee
The dynamical torsion with a constant background metric $g_{\mu\nu}$ yields:
\be
H^{t\theta\phi} = \frac{(2\pi\alpha')^{3/2}P }{r^4 \sin\theta } \ .\label{T-21}
\ee
It satisfies the form field equations: 
\be
\partial_{\lambda } \,  {H^{\lambda \mu \nu} } +  \frac{1}{2} \;(g^{\alpha \beta} \,   \partial_\lambda  \, {g_ {\alpha \beta}})\, H^{\lambda \mu \nu}=\ 0\ .\label{T-22}
\ee
A torsion incorporates a new (tensor) connection and modifies the covariant derivative: $\nabla_{\mu}\rightarrow {\cal D}_{\mu}$. A torsion connection has been worked out in a second order formalism by the authors \cite{abhishek-JHEP}. A torsion redefines a non-linear gauge theoretic curvature $F_{\mu\nu}$ which may be interpreted as a geometric one in the formalism (\ref{T-23}). The $A_{\mu}$ on a $3$-brane satisfies: ${\cal D}_{\mu} {\cal F}^{\mu \nu} =\ 0$. Explicitly, the field equations are:
\be
\partial_{\mu} {\cal F}^{\mu \nu}\ + \ \frac{1}{2} (g^{\alpha \beta}\ \partial_{\mu}\ g_{\alpha \beta})\ {\cal F}^{\mu \nu}\ - \ \frac{1}{2}\ 
{\cal H_{\alpha \beta}}^{\nu}\ {\cal F} ^{\alpha \beta} =\ 0\ .\label{T-24}
\ee 
The non-zero components of the geometric ${\cal F}_2$ curvature become
\bea
{\cal F}_{tr}&=&-\ \frac{P}{r^2} \left( Q_1 \sin\left [ \frac{P{\sqrt{2\pi\alpha'}}}{r}\right ]\ +\ Q_2\cos\left [\frac{P{\sqrt{2\pi\alpha'}}}{r}\right ]\right )\nonumber\\
{\rm and}\quad {\cal F}_{\theta\phi}&=& {{P}\over{2\pi\alpha'}}\left(Q_1 \cos \left[\frac{P{\sqrt{2\pi\alpha'}}}{r}\right] - Q_2 \sin \left[\frac{P{\sqrt{2\pi\alpha'}}}{r}\right]\right)\sin\t\ .\label{T-3}
\eea
A non-linear magnetic charge, underlying the non-trivial field component ${\cal F}_{\t\phi}$ in absence of a magnetic monopole, is remarkable. The new phenomenon may rule out a theoretical notion of a magnetic point charge. It ensures the significant role played by a torsion in the framework. In fact the observation is in agreement with the Kerr-Newman black hole where an angular momentum is seen to mix an electric field component with a magnetic field \cite{sunita-IJMPA}. An emergent quantum geometry on a vacuum created $3$-brane may primarily be influenced by a geometric torsion ${\cal H}_3$ and a gauge theoretic two form $F_2$. A priori two independent sources for a nontrivial curvature on a pair of $(3{\bar 3})$-brane may seen to possess their origin in an effective torsion curvature scalar ${\tilde{\cal K}}$ on a $D_4$-brane \cite{abhishek-JHEP}. The open string metric $G_{\mu\nu}^{(NS)}$  receives further corrections due to a geometric torsion in five dimensional brane world. The generic curvature scalar ${\tilde{\cal K}}$ on $S^1$ describes ${\cal H}_3$ in eq(\ref{T-2}) and ${\cal F}_2$ in eq(\ref{T-3}) on a $3$-brane. Then, the emergent metric on a gravitational $3$-brane \cite{abhishek-JHEP,abhishek-PRD,abhishek-NPB-P,sunita-NPB,sunita-IJMPA} is given by
\be
{G}_{\mu\nu}=\left ( G^{(NS)}_{\mu\nu}\ +\ C_1\ {\bar{\cal F}}_{\mu\alpha}\ g^{\alpha\beta}{\bar {\cal F}}_{\beta\nu}\ +\ 
C_2\ {\bar{\cal H}}_{\mu\alpha\beta}\ g^{\alpha\delta}g^{\beta\sigma}{\cal H}_{\delta\sigma\nu}\right )\ ,\label{T-4}
\ee
where $(C_1,C_2)$ are arbitrary constants. We set ($C_1=\pm 1$, $C_2=\pm 1/2$) and consider $(2\pi\alpha')=1$ for simplicity.

\subsection{Axionic sourced brane universe}
The non-perturbative quantum geometries, a priori on a vacuum created pair of gravitational $(3{\bar 3})$-brane are worked out using the metric (\ref{T-4}). The geometric patches are given by
\bea
G_{tt}&=&- \left(1 \mp \frac{P^2}{r^4} \pm \frac{P^2}{r^4}\left( Q_1 \sin\frac{P}{r}+ Q_2 \cos\frac{P}{r}\right)^2\right)\ ,\nonumber\\ 
G_{rr}&=& \left(1  \pm \frac{P^2}{r^4}\left( Q_1 \sin\frac{P}{r}+ Q_2 \cos\frac{P}{r}\right)^2\right)\ ,\nonumber\\
G_{\t\t}&=& r^2\left(1 \mp \frac{P^2}{r^4} \mp \frac{P^2}{r^4}\left( Q_1 \cos\frac{P}{r}- Q_2 \sin\frac{P}{r}\right)^2\right)\nonumber\\
G_{\phi\phi}&=&G_{\t\t}\ \sin^2\t \ .\label{T_5}
\eea
The emergent patches may be re-expressed with an interchange between $dr\longleftrightarrow rd\theta$. Under an interchange, the conserved charges rename appropriately and geometry may be given by 
\bea
ds^2&=&-\ \left(1 \mp \frac{P^2}{r^4} \pm \frac{P^2}{r^4}\left( Q_1 \sin\frac{P}{r}+ Q_2 \cos\frac{P}{r}\right)^2\right)dt^2\nonumber\\
&&+\ \left(1 \mp \frac{P^2}{r^4} \mp \frac{P^2}{r^4}\left( Q_1 \cos\frac{P}{r}- Q_2 \sin\frac{P}{r}\right)^2\right) dr^2\nonumber\\
&&+\ \left(1  \pm \frac{P^2}{r^4}\left( Q_1 \sin\frac{P}{r}+ Q_2 \cos\frac{P}{r}\right)^2\right)r^2d\theta^2\nonumber\\
&&+\ \left(1 \mp \frac{P^2}{r^4} \mp \frac{P^2}{r^4}\left( Q_1 \cos\frac{P}{r}- Q_2 \sin\frac{P}{r}\right)^2\right)r^2\sin^2\theta d\phi^2\ .\label{T-6}
\eea
For simplicity, we consider $Q_1=Q_2=Q$ in the paper. A priori, the quantum geometries on a gravitational $3$-brane reduce to yield:
\bea
ds^2&=&- \left(1 \mp \frac{P^2}{r^4} \pm \frac{P^2Q^2}{r^4}\left(1+\sin\frac{2P}{r}\right)\right)dt^2\ \nonumber \\
&& +\ \left(1 \mp \frac{P^2}{r^4} \mp \frac{P^2Q^2}{r^4}\left(1-\sin\frac{2P}{r}\right)\right) dr^2\nonumber\\
&&+\ \left(1  \pm \frac{P^2Q^2}{r^4}\left(1+\sin\frac{2P}{r}\right)\right)r^2d\theta^2\nonumber\\
&&+\ \left(1 \mp \frac{P^2}{r^4} \mp \frac{P^2Q^2}{r^4}\left(1-\sin\frac{2P}{r}\right)\right)r^2\sin^2\theta d\phi^2\ .\label{T-8}
\eea
A global scenario is defined with a vacuum created gravitational pair of $(3{\bar 3})$-brane from a BPS $D_4$-brane underlying a two form $U(1)$ gauge theory. The momentum conservation of a pair at its creation may intuitively imply a formal swing $r\rightarrow -r$ between the brane geometries and that on an anti-brane. It is interesting to note that a repulsive gravity sourced by a Newtonian like potential would like to drive away a brane universe away from an anti-brane. The interpretation provokes thought to believe the role of a quintessence in the string-brane model. The brane geometries are
approximated in a brane window: $r>|P|$ and $r>|Q|$ with $r^8>>(P^4,Q^4)$. Then the quantum geometries, sourced by a geometrical torsion on a vacuum created pair of $(3{\bar 3})$-brane, become
\bea
ds^2&=&- \left(1 \mp \frac{P^2}{r^4} \pm \frac{P^2Q^2}{r^4}\left(1+\sin\frac{2P}{r}\right)\right)dt^2\ \nonumber \\
&& +\ \left(1 \pm \frac{P^2}{r^4} \pm \frac{P^2Q^2}{r^4}\left(1-\sin\frac{2P}{r}\right)\right)^{-1} dr^2\nonumber\\
&&+\ \left(1  \pm \frac{P^2Q^2}{r^4}\left(1+\sin\frac{2P}{r}\right)\right)r^2d\theta^2\nonumber\\
&&+\ \left(1 \mp \frac{P^2}{r^4} \mp \frac{P^2Q^2}{r^4}\left(1-\sin\frac{2P}{r}\right)\right)r^2\sin^2\theta d\phi^2\ .\label{T-89}
\eea
It is remarkable to note that the quantum geometries in a low energy limit, $i.e.$ for a non-propagating torsion, reduces to Minkowski space-time. This fact makes the quantum geometries special as Einstein vacuum cannot be realized from the quantum vacua. It is indeed a property of the free string theory and hence corresponds to an exact theory of quantum gravity underlying a non-interacting graviton. Analysis reveals that the vacuum created pair of $(3{\bar 3})$-brane, underlying the quantum vacua (\ref{T-89}), are non-gravitational in nature!
Explicitly, the emergent quantum geometries reduce to yield:
\bea
ds^2_1&=&-\ \left(1 + \frac{M_Q+M_T}{r^4}\right)dt^2\ +\ \left(1 + \frac{M_Q-M_T}{r^4}\right)^{-1}dr^2\ +\left(1  + \frac{M_Q}{r^4}\right)r^2d\theta^2\ \nonumber\\
&& +\ \left(1 - \frac{M_Q-M_T}{r^4}\right)r^2\sin^2\theta d\phi^2\ .\label{T-91}
\eea
\bea
ds^2_2&=&-\ \left(1 - \frac{M_Q-M_T}{r^4}\right)dt^2\ +\ \left(1 - \frac{M_Q+M_T}{r^4}\right)^{-1}dr^2\ +\ \left(1  - \frac{M_Q}{r^4}\right)r^2d\theta^2\ \nonumber\\
&&  +\ \left(1 + \frac{M_Q+M_T}{r^4}\right)r^2\sin^2\theta d\phi^2\ .\label{T-92} 
\eea
\bea
ds^2_3&=&-\ \left(1 + \frac{M_Q-M_T}{r^4}\right)dt^2\ +\ \left(1 + \frac{M_Q+M_T}{r^4}\right)^{-1}dr^2\ +\ \left(1  + \frac{M_Q}{r^4}\right)r^2d\theta^2 \nonumber\\
&&  +\ \left(1 - \frac{M_Q+M_T}{r^4}\right)r^2\sin^2\theta d\phi^2\ .\label{T-93}
\eea
\bea
ds^2_4&=&-\ \left(1 - \frac{M_Q + M_T}{r^4}\right)dt^2\ +\ \left(1 - \frac{M_Q-M_T}{r^4}\right)^{-1}dr^2\ +\ \left(1  - \frac{M_Q}{r^4}\right)r^2d\theta^2 \nonumber\\
&& +\ \left(1 + \frac{M_Q - M_T}{r^4}\right)r^2\sin^2\theta d\phi^2\ ,\label{T-94}
\eea
where $M_Q=(2\pi\alpha')^{-1/2}(PQ)^2$ and $M_T=(2\pi\alpha')^{-1/2}P^2$ may define the microscopic black hole mass on a vacuum created pair of $(3{\bar 3})$-brane. The emergent geometries are asymptotically flat. They are characterized by two independent conserved charges $(P,Q)$. Firstly a torsion charge $P$ re-assures the significance of string charge to the quantum geometries in the formalism. Secondly a non-linear electric charge $Q$ may seen to incorporate a degeneracy in the quantum geometries (\ref{T-91}) and (\ref{T-93}) defined with $Q=0$. Needless to mention that two independent non-trivial universes began on a pair of $(3{\bar 3})$-brane at an event horizon by the KR flux in the $U(1)$ gauge theory on a $D_4$-brane. Subsequently, a lower form, $i.e.$ an one form flux, in disguise of a non-linear axion on a vacuum created $3$-brane within a pair creates a degenerate brane universes \cite{sunita-IJMPA}. In principle, a higher form, dual to the gauge field, on an appropriate $D_p$-brane for $p>3$ may possess a potential strength to vacuum create a large degeneracy and hence may be described by various clones of a semi-classical brane universe.

\sp
\noindent
Three torsion geometries, among the four (\ref{T-91})-(\ref{T-94}), are characterized by an event horizon and hence they describe various brane universes in the quantum regime. We shall see that the emergent geometry (\ref{T-93}) does not describe a brane universe as the curvature singularity described in a fundamental theory is not protected by an event horizon. Hence the brane universe is not allowed by a cosmic censorship hypothesis. The Ricci scalar ${\cal R}$ blows up at various values in $r$ in the vacuum created brane universe(s).

\sp
\noindent
In addition to a curvature singularity at $r\rightarrow 0$ in all four geometries, they are also described by ${\cal R}\rightarrow \infty$ when: (i) $r^2\rightarrow P{\sqrt{1+Q^2}}$ in eq(\ref{T-91}), (ii)  $r^2\rightarrow P{\sqrt{Q^2-1}}$ and $r^2\rightarrow (PQ)$ in eq(\ref{T-92}), (iii) $r^2\rightarrow P{\sqrt{1-Q^2}}$ in eq(\ref{T-93}) and (iv) $r^2\rightarrow (PQ)$ and $r^2\rightarrow P{\sqrt{1-Q^2}}$ in eq(\ref{T-94}). The lower bounds on $r>(P,Q)$ rule out the singularity at $r\rightarrow 0$ in a brane universe. In absence of an upper bound on $Q$, $i.e.\ Q^2>2$, all the created brane universes (\ref{T-92})-(\ref{T-94}) appear to possess a single curvature scalar ${\cal R}$ singularity at $r^2\rightarrow (PQ)$. Presumably they are described in an effective type IIA or IIB superstring theory on $S^1$ underlying a $D_4$-brane. Nevertheless the quantum black holes underlying the brane universes are described by an effective torsion curvature scalar:
\be
{\cal K}= \ {{3M_T}\over{2r^4}} \ .\label{T-941}
\ee
It shows that a brane universe does not access to a curvature singularity at $r\rightarrow 0$ which is otherwise present in a fundamental theory of gravity. The curvature underlying a varying energy density has been argued to be sourced by a quintessence axion \cite{priyabrat-EPJC} or an anti quintessence axion \cite{priyabrat-IJMPA} in the string-brane model. A varying positive energy density in the quantum regime presumably ensures an underlying de Sitter at the Big Bang.

\section{Non-perturbative correction}
\subsection{Quantum black hole}
The vacuum created pair of gravitational brane universes may be approximated by a Schwarzschild black hole in presence of a correction underlying a flat metric. Formally, the ${\cal F}_2$ contribution may be separated out from the ${\cal H}_3$. The emergent geometries (\ref{T-92}) and (\ref{T-94}) may be given by
\bea
ds^2&=&-\ \left(1 - \frac{M_Q}{r^4} \right)dt^2\ +\ \left(1 - \frac{M_Q}{r^4}\right)^{-1} dr^2
\ +\ \left(1 - \frac{M_Q}{r^4}\right)r^2d\theta^2\nonumber\\
&&+\ \left(1 + \frac{M_Q}{r^4}\right)r^2\sin^2\theta d\phi^2\ \mp\ \frac{M_T}{r^4}\left(-dt^2 +dr^2+r^2 \sin^2\theta d\phi^2\right)\ .\label{T-10}
\eea
The second term with a conformal factor ($M_T/r^4$) couples to a trivial metric and may formally be identified with a membrane world-volume in the formalism. The RR charge with a $D$-membrane ensures a non-perturbative nature \cite{polchinski} underlying the quantum correction to a peturbative (low energy) string vacuum. Thus the lower dimensional $D$-brane correction term incorporates quantum effects non-perturbatively into the otherwise semi-classical vacuum. 

\sp
\noindent
In a low energy limit, the non-perturbative correction decouples to yield a semi-classical Schwarzschild geometry on a vacuum pair of $(3{\bar 3})$-brane. The reduced black hole is characterized by an event horizon at $r\rightarrow r_e=(M_Q)^{1/4}={\sqrt{PQ}}$. An event horizon, described by a non-linear $U(1)$ charge in a gauge theory, is known to describe an effective black hole on a $3$-brane. It implies that the event horizon is essentially described by a non-linear conserved charge $(PQ)$ underlying a geometric torsion in the string-brane model. 

\sp
\noindent
Formally the charge $M_Q$ is identified with a mass of a $4D$ black hole obtained on a brane universe. 
The mass appears to break the spherical symmetry in the brane universe. The deformation 
geometry may also be realized intuitively with the help of an extra fifth hidden dimension transverse to a $D_3$-brane or an anti $D_3$-brane. A large extra dimension brings in some of the low energy closed string modes into a vacuum created $3$-brane or an anti $3$-brane. Thus a geometric torsion  in the formalism arguably incorporates non-perturbative quantum effects, via $D$-branes, to the low energy string effective action.

\sp
\noindent
The empirical formula underlying an effective gravitational potential, sourced by a geometric torsion, hints at three extra dimensions in the gauge choice. They may further be analyzed: (i) with a medium energy laboratory scale underlying a large radial coordinate: $r^2=\rho>0$ and (ii) in a low energy scale underlying a very large radial coordinate: $r^4=R>0$ for a classical black hole presumably in Einstein vacuum.

\subsection{Laboratory black hole}
An effective microscopic black hole on a brane universe at a Planck scale may reduce to a laboratory black hole defined in a medium energy range.
With a larger scale, $i.e.\ \rho=r^2>0$, an effective Schwarzschild may approach a Schwarzschild black hole in Einstein vacuum. With a rescaling, the brane universe in a medium energy regime may be approximated by
\bea
ds^2&=&-\ 4\rho\left(1 - \frac{M_Q}{\rho^2} \right)dt^2\ +\ \left(1 - \frac{M_Q}{\rho^2}\right)^{-1} d\rho^2\nonumber\\
&&\ +\ 4\left(1 - \frac{M_Q}{\rho^2}\right)\rho^2 d\theta^2\ +\ 4\left(1 + \frac{M_Q}{\rho^2}\right)\rho^2\sin^2\theta d\phi^2\nonumber\\
&&\ \mp \frac{M_T}{\rho^2}\left(-4\rho dt^2 + d\rho^2 + 4\rho^2 \sin^2\theta d\phi^2\right)\ .\label{T-111}
\eea
Some of the geometric patches may seen to decouple from the emergent geometry in the regime. Then the effective brane universe may be approximated by
a quantum Schwarzschild black hole a priori with an effective $D$-string world-volume correction in the framework. It is given by
\bea
&&ds^2 =\ ds^2_I\ +\ ds^2_{II}\ ,\nonumber\\
{\rm where}&&ds^2_I=-\ 4\rho\left(1 - \frac{M_Q}{\rho^2} \right)dt^2\ +\ \left(1 - \frac{M_Q}{\rho^2}\right)^{-1} d\rho^2\ +\ 4\rho^2 d\Omega^2\nonumber\\
{\rm and}&&ds^2_{II}=\ \mp\ \frac{M_T}{\rho^2}\left (-4\rho\ dt^2 + d\rho^2\right)\ .\label{T-112}
\eea
The effective gravitational potential hints at a large extra dimension in a brane universe. In fact it is in conformity with an underlying $D_4$-brane world-volume. However the $D$-string correction $dS^2_{II}$ is primarily associated with a torsion or an axionic charge $M$ on a pair of $(3{\bar 3})$-brane. Unlike to a flat metric on a membrane in eq(\ref{T-10}), the string world volume appears to possess a nontrivial Ricci curvature scalar ${\cal R}_{II}=1/2\rho^2$. 

\sp
\noindent
Nevertheless, the curvature underlying a correction is evaluated at the event horizon of a semi-classical black hole to yield a positive constant vacuum energy density. The scalar ${\cal R}_I$ for the line-element $ds^2_I$ may be checked to yield a positive constant energy density at an event horizon, $i.e.\ \rho\rightarrow \rho_e=P{\sqrt{2}}$. 
In a low energy limit the non-perturbative geometric correction becomes insignificant and the effective brane universe may describe a spherically symmetric $4D$ Schwarzschild black hole in presence of a large extra dimension. 

\subsection{Semi-classical brane universe}
A primordial black hole underlying a brane universe may be obtained from the quantum geometries for a relatively large length scale. It may be approximated by a re-defined large radial coordinate with $\ r^4=R>0$. In the case, the quantum brane geometries may formally approach to a Schwarzschild black hole after decoupling an axionic charge. In the regime, the brane geometries may be approximated by
\bea
ds^2&=&-\ 16R^{3/2}\left(1 - \frac{M_Q}{R} \right)dt^2\ +\ \left(1 - \frac{M_Q}{R}\right)^{-1} dR^2\ \nonumber\\
&& + \ 16 \left(1 - \frac{M_Q}{R}\right){R^2} d\theta^2+ 16 \left (1 + \frac{M_Q}{R}\right ) R^2 \sin^2\theta d\phi^2\ \nonumber \\ && \mp\ \frac{M_T}{R}\left (-16 R^{3/2}\ dt^2 + dR^2 + 16R^2 \sin^2\theta d\phi^2\right)\ \label{T-121}
\eea 
A non-perturbative correction underlying an axionic charge $M_T$, with a very large $R$ scale, may seen to describe a varying negative curvature scalar 
${\cal R}={{-9}\over{8R^2}}$ on a vacuum created gravitational pair of $(3{\bar 3})$-brane. The correction in energy density is computed at the Schwazchild event horizon $R\rightarrow R_e=M_Q$ to confirm a small negative energy vacuum density. Then the energy density in a Schwarzschild like 
black hole is worked out at the horizon to confirm a positive vacuum energy density. The Ricci scalar blows up at the event horizon $R_e$. Then the semi-classical geometry in a brane universe is approximated to yield:
\bea
ds^2&=& -\ 16R^{3/2}\left(1 - \frac{M_Q}{R} \right)dt^2\ +\ \left(1 - \frac{M_Q}{R}\right)^{-1} dR^2\  \nonumber\\
&& +\ 16\left ( 1 - {{M_Q}\over{R}}\right ) R^2 d\t^2 \ + \ 16 \left ( 1 + {{M_Q}\over{R}}\right ) R^2 \sin^2\t d\phi^2\ .\label{T-122}
\eea
It implies that the spherical symmetry is broken in a semi-classical black hole which is described by a torsion-less 
vacuum in the formalism. Unlike to a Schwarzschild black hole in Einstein vacuum, the Schwarschild brane universe is not Ricci flat. A computation of Ricci curvature scalar ${\cal R}$ in the semi-classical brane universe ensures naive singularities at $R\rightarrow 0$ and $R \rightarrow M$. However they are not accessed by the brane universe as the large radial coordinate is bounded below by $(P,Q)<R$. Importantly, the curvature scalar 
turns out to be negative on a brane window. The varying curvature presumably ensures an underlying quintessence AdS black hole \cite{pssk} in the string-brane model. At an event horizon, the vacuum energy density takes a negative constant value to re-assure an AdS brane geometry there. Thus in a low energy limit, a brane universe on a vacuum created gravitational pair of $(3{\bar 3})$-brane is presumably described by an asymptotic AdS in Einstein vacuum. In fact, an asymptotic analysis for the near horizon Schwarzschild brane universe in ref \cite{abhishek-JHEP} is in agreement with the obtained result.

\subsection{Torsion decoupling regime}
A torsion sourced by a string charge decouples from the semi-classical brane universe in a low energy limit. It may as well be described by a large $r$ where a propagating torsion freezes and lead to a condensate. It has been argued that the geometries, right after their vacuum creation at a cosmological horizon, undergo quantum tunneling and may be viewed through radiations \cite{abhishek-JHEP}. Since an observer in a brane window is not accessible to a curvature singularity otherwise present in Einstein gravity, tunnelling aspects of a vacuum to another becomes significant in the quantum regime. It has been argued that an effective de Sitter universe cools down via a series of geometric transitions in the early epoch. Finally it lead to a semi-classical universe in $(3+1)$-dimensions. Thus, a low energy brane universe may well be approximated by the Ricci scalar and hence the curvature singularity becomes significant there. In fact, we shall see that the brane universes (\ref{T-91}), (\ref{T-92}) and (\ref{T-94}), on a gravitational pair of vacuum created $(3{\bar 3})$-brane, tend to approach a single vacuum in Einstein gravity. Analysis reveals that two of the brane universes (\ref{T-91}) and (\ref{T-93}) may causally identify with the remaining two brane geometries (\ref{T-92}) and (\ref{T-94}) respectively across an event horizon where a light cone flips by a right angle. 

\sp
\noindent
In a low energy limit, the emergent quantum geometries dissociate an higher order term in $(1/r)$ defined via an effective gravitational potential purely sourced by a two form flux. The reduced semi-classical geometries in the limit may be approximated by the torsion-less effective brane universes. A priori the degenerate quantum vacua (\ref{T-91})-({\ref{T-94}) reduce to yield:
\bea
ds^2&=&-\ \left(1 + \frac{M_Q}{r^4}\right)dt^2\ +\ \left(1 + \frac{M_Q}{r^4}\right)^{-1}dr^2\ 
+\ \left(1  + \frac{M_Q}{r^4}\right)r^2d\theta^2\nonumber\\
&&+\ \left(1 - \frac{M_Q}{r^4}\right)r^2\sin^2\theta d\phi^2\ .\label{T-95}
\eea
\bea
ds^2&=&-\ \left(1 - \frac{M_Q}{r^4}\right)dt^2\ +\ \left(1 - \frac{M_Q}{r^4}\right)^{-1}dr^2\ 
+\ \left(1  - \frac{M_Q} {r^4}\right ) r^2d\theta^2\nonumber\\
&&+\ \left(1 + \frac{M_Q}{r^4}\right)r^2\sin^2\theta d\phi^2\ .\label{T-96}
\eea
Under a flip of light cone across an event horizon, when considered with an interchange of angular coordinates, the brane universe (\ref{T-96}) identifies with the other brane geometry (\ref{T-95}). Formally they may be identified under an interchange: $M_Q \leftrightarrow -M_Q$. 

\section{Quintessence}
A quintessence axion, presumably on a vacuum created ${\bar 3}$-brane, is argued to source an extra fifth dimension to our brane universe on a $3$-brane with a vacuum pair. The curvature scalar ${\cal{K}}$ on a gravitational anti $3$-brane may describe a quintessence axion and a geometric ${\cal F}_2$ governs the torsion curvature on a gravitational $3$-brane. An accelerated and expanding universe in the present day cosmology may seen to be influenced by 
a growth in extra dimension between a brane and an anti-brane universes. The high energy torsion modes have been argued to decouple with a large fifth dimension and eventually a low energy brane universe may appear to decouple from the anti-brane in the limit. The decoupling of non-perturbative quantum effects may lead to describe Einstein vacuum in the formalism.

\sp
\noindent
In this section we perform Weyl scaling(s) in the effective degenerate vacua (\ref{T-91})-(\ref{T-94}) and analyze the de Sitter and AdS brane geometries in the framework to signal the presence of a quintessence. Generically a metric under a Weyl scaling in a low energy effective string action describes Einstein gravity without any change in its causal structure. However a Weyl scaling of the effective metric on a $D$-brane in the framework may seen to scale the vacuum with a higher or a lower energy density \cite{abhishek-JHEP,abhishek-PRD,priyabrat-EPJC,priyabrat-IJMPA}. This is due to a fact that the emergent geometries are defined on a brane window with a bound on its world-volume coordinate $r$. 

\subsection{de Sitter brane}
We obtain effective de Sitter geometries by performing a conformal Weyl transformation among the effective geometric patches approximated  by the line-elements (\ref{T-91})-(\ref{T-94}) on a gravitational pair $(3{\bar 3})$-brane. A scale factor $\lambda=r^2$ is worked out to obtain a de Sitter brane vacua in this section. Under a scaling the effective brane universe may be approximated by
\be
ds^2=\ -\; {1\over{r^2}}dt^2\  +\ \left ( r^2 - \frac{M_Q-M_T}{r^2}\right )^{-1} dr^2\ +\ d\Omega^2\label{qads-71}
\ee
\be
{\rm and}\qquad ds^2=\ -\; {1\over{r^2}} dt^2\  +\ \left ( r^2 - \frac{M_T + M_Q}{r^2}\right )^{-1} dr^2\ +\  d\Omega^2\ .\quad\quad\;\;\ {} \label{qads-72} 
\ee
It may be recalled that a NS two form in a string world-sheet is known to incorporate a string charge which may be viewed as a torsion charge in an effective string theory. Thus, a string charge may be identified with an axionic charge on a $3$-brane. The brane universe(s) in the quantum phase 
may imply a decoupling of $S_2$-symmetric patch from the causal patches and hence may be viewed through a pair of lower dimensional brane \cite{abhishek-JHEP}. The curvature scalars ${\cal R}_{\pm}$ computed respectively for the black holes mass $(M_Q\pm M_T)$. They are:
\be
{\cal R}_{\pm}=\ {{6}\over{r^4}} \left (M_Q\pm M_T\right )\ .\label{qads-8}
\ee
A positive curvature scalar in the quantum regime ensures a de Sitter in presence of a non-perturbative correction underlying a $D$-brane in 
the ten dimensional type IIA and type IIB superstring theories on $S^1$. At this point we recall that a superstring theory describes an AdS vacuum.
A positive vacuum energy density, underlying a de Sitter vacuum, in the framework is sourced by a non-linear electric field associated with a $D$-brane correction to the $AdS$ string vacuum. Formally our analysis leading to a de Sitter vacuum in superstring theory is 
in conformity with the idea of de Sitter vacua remarkably constructed in presence of $D$-branes in string theory in a different context \cite{kklt}. Most importantly a varying positive curvature may seen to be sourced by a quintessence axion underlying a repulsive gravity between and brane and an anti brane. Nevertheless, the vacuum energy density at the event horizon signifies a small positive cosmological constant in Einstein gravity.

\subsection{AdS brane}
We perform a conformal Weyl transformation in the effective degenerate geometries obtained (\ref{T-91})-(\ref{T-94}) on vacuum created gravitational pair of $(3{\bar 3})$-brane. A scale factor $\lambda=1/r^2$, predefined with $r^2>(M_Q,M_T)$, is worked out to obtain an AdS brane vacuum in this section. Under a scaling the effective brane universe may a priori be described by
\be
ds^2=-\left ( r^2 + \frac{M_T \pm M_Q}{r^2}\right )dt^2\  +\ \left (r^2 - \frac{M_T\mp M_Q}{r^2}\right )^{-1} r^4 dr^2\ 
+\ r^4 d\Omega^2\label{qads-1}
\ee
\bea
{\rm and}\qquad ds^2&=&\ -\left(r^2 - \frac{M_T \mp M_Q}{r^2}\right)dt^2 + \left(r^2 + \frac{M_T \pm M_Q}{r^2}\right)^{-1} r^4dr^2\nonumber \\ &&\ +\ r^4\ d\Omega^2\ ,\label{qads-2} 
\eea
where the brane geometries are approximated in a window for $r^4>>(M_T^2, M_Q^2)$. However the causal patches sourced by an axion appears to be associated with a relative wrong sign. The relative sign in $G_{tt}$ and $G_{rr}$ may be corrected via a matrix projection underlying a discrete
transformation among the causal patches \cite{abhishek-JHEP,abhishek-PRD}. The causal patches define a $(2\times 2)$  matrix and it is given by
\begin{equation}
{\cal M}=\frac{1}{2{\sqrt{M_Q}}}\left( \begin{array}{ccc}
{G}_{tt}(+)&& {G}_{rr}(+)\\
&& \\
{G}_{rr}(-)&&{G}_{tt}(-)
\end{array} \right)\ , \label{qads-3}
\end{equation}
where
\bea
{G}_{tt}(+)\ =\ -\left( r^2 + \frac{M_T}{r^2}  \pm \frac{M_Q}{r^2}\right)\ ,\; \; 
{G}_{rr}(+)\ =\ \left ( r^2 + \frac{M_T}{r^2} \pm \frac{M_Q}{r^2}\right )^{-1}\ ,\nonumber\\
{G}_{tt}(-)\ =\ -\left ( r^2 - \frac{M_T}{r^2}  \pm \frac{M_Q}{r^2}\right )\ ,\;\; {G}_{rr}(-)\ =\ \left ( r^2 - \frac{M_T}{r^2} \pm \frac{M_Q}{r^2}\right )^{-1}\ .\label{qads-4}
\eea
The causal patches in the quantum geometries (\ref{qads-1})-(\ref{qads-2}) 
may be viewed via a projection of the matrix ${\cal M}$ on the column vectors: 
\bea
\left( \begin{array}{c}
1\\
\\  
0
\end{array}\right)\qquad {\rm and}\qquad \left( \begin{array}{c}
0\\
\\
1
\end{array}\right)
\ . \nonumber
\eea
The matrix determinant turns out to be independent of a lower form non-linear charge $Q$ or $M_Q$ and is given by: $\det {\cal M}=-1$. It may imply a discrete transformation of the geometric causal patches in an effective quantum gravity on a brane universe. The inverse of the matrix becomes
\be
{\cal M}^{-1}=\frac{-1}{2{\sqrt{M_Q}}}\left( \begin{array}{ccc}
G_{tt}(-) & & -{G}_{rr}(+)\\
 && \\
-{G}_{rr}(-)&&{G}_{tt}(+)
\end{array} \right )\ .\label{qads-5}
\ee
An inverse matrix projection on the column vectors may be checked to describe the quantum geometries with appropriate causal patches. In other words the
inverse matrix projected geometries are obtained by a discrete transformation from the matrix projected ones in eqs(\ref{qads-1}) and (\ref{qads-2}). A  discrete transformation may seen to reduce the degeneracy in the quantum vacua by fixing a $(-)$-sign in the effective metric potential sourced by a charge $M_Q$. Two, among the four emergent, geometries may seen to describe a brane universe. Under an analytic continuation to real time they may be given by
\be
ds^2=-\left ( r^2 - \frac{M_Q-M_T}{r^2}\right )dt^2\  +\ \left (r^2 - \frac{M_Q-M_T}{r^2}\right )^{-1} r^4 dr^2\ 
+\ r^4 d\Omega^2\label{qads-61}
\ee
\bea
{\rm and}\quad ds^2=-\left(r^2 - \frac{M_Q + M_T}{r^2}\right)dt^2  + \left( r^2 - \frac{M_Q+M_T}{r^2}\right)^{-1} r^4dr^2  + r^4 d\Omega^2\ .\ \label{qads-62} 
\eea
The curvature scalars ${\cal R}_{\pm}$ respectively for the effective black holes with mass: ($M_Q\pm M_T$) are computed to confirm an AdS brane with varying curvature. Explicitly they may be given by
\be
{\cal R}_{\pm}= -\;  {{6}\over{r^4}}\left ( {{M_Q\pm M_T}\over{r^4}}\ +\ 2\right )\ .\label{qads-611}
\ee
The varying curvature presumably imply an accelerating brane universe and hence hint for a quintessence axion. At an event horizon the curvature tends to a constant and precisely describes an AdS brane underlying a constant vacuum energy density. In the low energy limit a string charge on a ${\bar D}_3$-brane may be drained off by a gauge theoretic charge. Thus for $M_Q=M_T$ one arrives at a single brane universe and the reduced geometry becomes
\be
ds^2=-\left ( r^2 - \frac{2M_Q}{r^2}\right )dt^2\  +\ \left ( r^2 - \frac{2M_Q}{r^2}\right )^{-1} r^4dr^2\ +\ r^4 d\Omega^2\ . {}\label{qads-612} 
\ee
In the case the cancellation of a torsional effects by a string charge removes the degeneracy in the vacuum solutions. Then the brane universe is described with a torsion free geometry. It may be identified with a semi-classical vacuum in Einstein gravity.
\subsection{Dark energy}
The effective geometries (\ref{qads-1}) and (\ref{qads-2}) for a large scale may further be analyzed for $r^2=\rho$ in the string-brane model.
The non-perturbative quantum effects apparently described by the four vacua are worked out with euclidean time. Explicitly, they are given by
\be
ds^2=\left ( \rho + \frac{M_T + M_Q}{\rho}\right )dt^2_e\  +\ {1\over{4}}\left (1 - \frac{M_T- M_Q}{\rho^2}\right )^{-1} d\rho^2\ 
+\ \rho^2 d\Omega^2\ ,\label{dark-1}
\ee
\be
ds^2=\left ( \rho + \frac{M_T - M_Q}{\rho}\right )dt^2_e\  +\ {1\over{4}}\left (1 - \frac{M_T + M_Q}{\rho^2}\right )^{-1} d\rho^2\ 
+\ 4 \rho^2 d\Omega^2\ ,\label{dark-2}
\ee
\be
ds^2={1\over{4}}\left ( 1 - \frac{M_T + M_Q}{\rho^2}\right )dt^2_e\  +\ \left ( \rho + \frac{M_T-M_Q}{\rho}\right )^{-1} \rho^2 d\rho^2\ +\ \rho^2d\Omega^2\ \label{dark-3} 
\ee
\be
{\rm and}\;\; ds^2={1\over{4}}\left(1 - \frac{M_T - M_Q}{\rho^2}\right)dt^2_e  + \left (\rho + \frac{M_T + M_Q}{\rho}\right)^{-1} \rho^2 d\rho^2 + \rho^2d\Omega^2\ .\label{dark-4} 
\ee
The vacua (\ref{dark-1})-(\ref{dark-3}) 
among the four geometric patches correspond to an AdS brane universe. Nevertheless, the emergent geometries (\ref{dark-3}) and (\ref{dark-4}) under an interchange $dt_e\leftrightarrow d\rho$ may seen to define a de Sitter brane universe. The mixed causal patches in the brane geometries (\ref{dark-1})-(\ref{dark-2}) and in the remaining geometries (\ref{dark-3})-(\ref{dark-4}) with $dt_e\leftrightarrow d\rho$ may be separated out using matrix ${\cal M}$ projection(s) (\ref{qads-3}) defined with appropriate normalization(s). The det${\cal M}=-1$ at an event horizon of a primordial 
black hole ensures a discrete transformation underlying the quantum patches. Three independent matrix projections on the ($2\times 1$) column vectors are worked out between the geometries: (\ref{dark-1})-(\ref{dark-2}), (\ref{dark-2})-(\ref{dark-3}) and (\ref{dark-1})-(\ref{dark-3}). The inverse matrix ${\cal M}^{-1}$ projections on the column vectors are checked to yield six redefined vacua. However 
two among the six vacua may seen to be defined with a correct relative sign in their effective potentials. With lorentzian signature they are: 
\be
ds^2=-\ \left ( 1 - \frac{M_Q - M_T}{\rho^2}\right )dt^2\  +\ \left ( 1 - \frac{M_Q-M_T}{\rho^2}\right )^{-1} d\rho^2\ +\ \rho d\Omega^2\ \label{dark-5} 
\ee
\be
{\rm and}\;\; ds^2=- \left( 1 - \frac{M_T + M_Q}{\rho^2}\right)dt^2  + \left(1 - \frac{M_T + M_Q}{\rho^2}\right)^{-1} d\rho^2 + 4\rho^2d\Omega^2\ .\label{dark-6} 
\ee
The non-perturbative quantum vacua may further be re-expressed in term of the decoupled geometries. They are:
\bea
ds^2&=&-\ \left ( 1 - {{\rho}\over{b}}- \frac{M_Q - M_T}{\rho^2}\right )dt^2\  +\ \left ( 1 - {{\rho}\over{b}} - \frac{M_Q-M_T}{\rho^2}\right )^{-1} \quad\quad\;\ {}\nonumber\\
&&-\ {{\rho}\over{b}} \left ( dt^2 + d\rho^2 - b d\Omega^2\right )\label{dark-5} 
\eea
\bea
{\rm and}\;\; ds^2&=&- \left( 1 - {{\rho}\over{b}} - \frac{M_Q + M_T}{\rho^2}\right)dt^2  + \left( 1 - {{\rho}\over{b}} - 
\frac{M_T + M_Q}{\rho^2}\right)^{-1} d\rho^2\ \nonumber\\
&&  +\ 4\rho^2d\Omega^2\ -\ {{\rho}\over{b}}\left ( dt^2 + d\rho^2\right ) \ ,\quad\; {}\label{dark-6} 
\eea
where $b>0$ is a constant. A decoupled causal patch from the $S_2$-symmetric geometry in eq(\ref{dark-5}) may signify an early epoch in de Sitter \cite{abhishek-JHEP}. The curvature scalars ${\cal R}_-$ and ${\cal R}_+$ respectively for the black holes in eqs(\ref{dark-5}) and (\ref{dark-6}) are computed to yield:
\bea
{\cal R}_-={{6}\over{\rho}}\left ( {1\over{b}} - {1\over{4\rho}}\right )\qquad {\rm and}\qquad {\cal R}_+={{6}\over{\rho^4}}\left ( M_Q-M_T\right )\ .\label{dark-7}
\eea
The varying positive scalar curvatures underlying the semi-classical black holes presumably describe a quintessence de Sitter brane in the framework. 
At an event horizon they may seen to underlie a de Sitter in Einstein gravity. The non-perturbative brane corrections, underlying the quantum effects, to the black holes are checked for their curvature scalar. They ensure a positive curvature and the conformal factor further hints at a quintessence axion. 
Interestingly the effective black holes on a vacuum created gravitational pair of $(3{\bar 3})$-brane may formally be seen to be in agreement with a quintessence de Sitter black hole without a cosmological constant in five dimensions \cite{chen-pan-jing}. The state parameter is worked out to yield $\omega_q=-3/4$ for a quintessence de Sitter black hole on a brane (\ref{dark-6}) which turns out to be within the observed range ($-1<\omega_q<-1/3$) in cosmology.

\sp
\noindent
For a special case, $i.e.\ M_Q=M_T$, the semi-classical geometry (\ref{dark-6}) reduces to a Ricci flat and describes a Schwarzschild black. It implies that the energy density is not a constant ($\Lambda=0$) in the formalism. Nevertheless a quintessence axion incorporates a varying vacuum energy density into the brane universe(s). The absence of a cosmological constant in the low energy vacua may well be understood from the absence of a constant two form in the ansatz (\ref{T-1}). Thus a constant NS field incorporates a cosmological constant while a dynamical NS field appears to govern a varying vacuum energy density underlying a quintessence axion in the string-brane model. Since a $D$-instanton is sourced by a quitessence axionic scalar field in our model, it may provoke thought to believe that $D$-instanton is a potential candidate to the dark energy in our universe.

\section{Concluding remarks}
Primarily the research work has been shown to explore new quantum geometries in $(3+1)$-dimensions on a pair of $(3{\bar 3})$-brane underlying a type IIA or IIB superstring theory on $S^1$. In particular, an effective torsion curvature in a second order formalism underlying a non-linear $U(1)$ gauge theory on a $D_4$-brane was explored to investigate certain aspects of a quintessence axion in a vacuum created brane universe. It was argued that a quintessence dynamics underlie an effective torsion curvature on an anti $3$-brane, while a $3$-brane dynamics may be approximated by a DBI action. A five dimensional propagating torsion underlying an effective curvature scalar on a $D_4$-brane was further supported by a pair production mechanism in a non-linear gauge theory. It was argued that a vacuum created gravitational pair of $(3{\bar 3})$-brane with nontrivial geometries may be sourced by the KR quanta on a $D_4$-brane. Interestingly, a propagating geometric torsion was shown to generate a non-linear magnetic charge from an electric point charge. The phenomenon may rule out the possibility for a magnetic pole in the string-brane model.

\sp
\noindent
An extra (or hidden) fifth dimension transverse to both the world-volumes has been shown to play a significant role in the string-brane model. 
A large extra dimension was argued incorporate the low energy closed string vacuum between a gravitational pair of $(3{\bar 3})$-brane. In other words, the string-brane set-up underlying a geometric torsion may seen to describe a low energy perturbative closed string vacuum in presence of non-perturbative $D$-brane and anti $D$-brane corrections. In the context, a flat metric has been shown to be associated with the quantum corrections. They indeed reassure the absence of closed string modes on a $D$-brane.

\sp
\noindent
The quantum geometries were shown to degenerate in presence of a string charge in the string-brane model. The emergent vacua are analyzed for various non-perturbative quantum corrections underlying a geometric torsion. It was shown that a lower dimensional $D$-brane incorporates quantum effects to the semi-classical vacuum. The quantum effects in an effective gravity were shown to be sourced by a quintessence axion, which may provide a clue to unfold the origin of dark energy in our universe. Arguably the quintessence possesses it origin in a $D$-instanton, which is also sourced by an axionic scalar in the string-brane model.

\sp
\noindent
Furthermore the emergent brane/anti-brane geometries were Weyl scaled to obtain a de Sitter brane and an AdS brane in the regime. It was shown that the high energy geometric patches in the brane universe are sourced by a dynamical geometric torsion. In a moderately low energy limit, the torsion charge decouples from a brane universe and the vacuum may formally be identified with Einstein gravity. A Ricci curvature scalar underlying various brane universes further re-assured the vital role of a quintessence axion in the model. One of our result was shown to be in formal agreement with that in higher dimensional Einstein gravity described in presence of a quintessence scalar. It may suggest that our analysis underlying a torsion curvature may play a significant role to explain the accelerated expansion of our universe.

\section*{Acknowledgments}
Authors gratefully thank Ashoke Sen for various useful discussions.

\sp

\def\anp{Ann. of Phys.}
\def\cmp{Comm.Math.Phys.\ {}} {}
\def\prl{Phys.Rev.Lett.}
\def\prd#1{{Phys.Rev.} {\bf D#1}}
\def\jhep{JHEP\ {}}{}
\def\cqg{Class.\& Quant. Grav.}
\def\plb#1{{Phys. Lett.} {\bf B#1}}
\def\npb#1{{Nucl. Phys.} {\bf B#1}}
\def\mpl#1{{Mod. Phys. Lett} {\bf A#1}}
\def\ijmpa#1{{Int.J.Mod.Phys.} {\bf A#1}}
\def\ijmpd#1{{Int.J.Mod.Phys.} {\bf D#1}}
\def\mpla#1{{Mod.Phys.Lett.} {\bf A#1}}
\def\rmp#1{{Rev. Mod. Phys.} {\bf 68#1}}
\def\jaat{J.Astrophys.Aerosp.Technol.\ {}} {}
\def \epj#1{{Eur.Phys.J.} {\bf C#1}} 
\def \jcap{JCAP\ {}}{}

\end{document}